\documentclass[twocolumn,prl,superscriptaddress,showpacs]{revtex4}
\usepackage{epsfig}
\usepackage{times}
\usepackage{amsmath}

\begin{document}

\title{
Spectral properties of a
partially spin-polarized one-dimensional Hubbard/Luttinger
superfluid}

\author{Adrian E.~Feiguin}
\affiliation{Microsoft Station Q, University of California, Santa Barbara, California 93106, USA}
\author{David~A.~Huse}
\affiliation{Department of Physics, Princeton University, New Jersey 08544, USA}

\date{\today}

\begin{abstract}
We calculate the excitation spectra of a spin-polarized Hubbard
chain away from half-filling, using a high-precision momentum-resolved time-dependent Density Matrix
Renormalization Group method.  Focusing on the 
$U<0$ case, we present in some detail the 
single-fermion, pair, density and spin spectra, and 
discuss how spin-charge separation is altered for this system.  
The pair spectra show
a quasi-condensate 
at a nonzero momentum proportional to the polarization, as expected for this 
Fulde-Ferrel-Larkin-Ovchinnikov-like superfluid.
\end{abstract}

\maketitle

Systems of interacting fermions obeying Fermi liquid theory exhibit a
one-to-one correspondence between their low-energy quasiparticle
excitations 
and those of a non-interacting Fermi gas.  The quasiparticles
have renormalized energy and spectral weight, but possess 
the same charge and spin quantum numbers as the corresponding
noninteracting
fermions.  
This scenario breaks down in one dimension (1D): in this case,
each Fermi surface reduces to two points in momentum space, at
$k=\pm k_F$.  The resulting Fermi-surface nesting is present at
all densities and spin polarizations and destabilizes the Fermi
liquid, converting it instead to
a Luttinger liquid \cite{luttinger,voit,giamarchi},  
even for a weak interaction. 

In a Luttinger liquid with zero spin polarization, the elementary
excitations are collective density fluctuations that carry only
either spin (``spinons") or charge (``holons").  These excitations
have different dispersions, and, obviously, do not carry the same
quantum numbers as the original ``bare'' fermions. This leads to the
spin-charge separation picture, in which a fermion injected into
the system separates (``fractionates'') into an (anti-)holon and a
spinon, each of them carrying a share of the fermion's quantum
numbers.  The phenomenon of spin-charge separation, and
fractionation of particles more generally, is an
important concept in strongly-correlated systems, and has
intrigued physicists for decades. Its signatures have been
observed experimentally in 1D organic conductors \cite{ttf},
metallic wires \cite{wires}, carbon nanotubes \cite{nanotubes},
and nanowires in semiconducting heterostructures
\cite{semiconductors}. Proposals have been made to seek for
evidence of these phenomena in cold
atomic gases 
\cite{recati,kecke}.

In the 1D Hubbard model, 
the low-energy spin and charge modes of the Luttinger liquid
decouple as long as the system either is at half-filling or has
zero spin-polarization.  However, if the system is away from
half-filling and has a nonzero magnetization, the collective modes
that constitute the elementary low-lying excitations 
are linear combinations of the spin and charge fields
\cite{kollath3,penc and solyom,vekua}, so although one still has a
Luttinger liquid with fractionalized fermions, 
it is no longer strictly a ``spin-charge separation'' scenario. 
The field-theoretical formulation of the Luttinger liquid theory
has proven very effective in describing the low-energy physics of
a variety of models.  However, a fully quantitative and general 
picture of how the spin and charge degrees of
freedom couple to form full-fledged fermions is still missing.

In this work, we will focus our attention on the negative-$U$
(attractive) 1D Hubbard model, away from half-filling and at
nonzero spin polarization.  This model can now be studied experimentally
with ultracold atoms in an optical lattice.  As usual, the Hamiltonian is
\begin{eqnarray}
H =  & - &  t\sum\limits_{i, {\sigma}} \left(c^\dagger_{i\sigma}
c_{i+1\sigma}+h.c.\right) + U \sum\limits_{i} n_{i\uparrow}
n_{i\downarrow} ~, \label{one}
\end{eqnarray}
where $c^\dagger_{\ell\sigma}$ creates a fermion with spin
$\sigma=\,\uparrow ,\downarrow$ at site $\ell$;
$n_{\ell\sigma}=c^\dagger_{\ell\sigma}c_{\ell \sigma}$;
$t$ is the hopping matrix element, which we set to unity
(we also set the lattice spacing to unity); and $U$ is the
interaction strength that in this work will be considered negative
(attractive). The negative and positive $U$ versions of this model
can be mapped exactly onto each other by the ``canonical''
transformation that applies a particle-hole and momentum change to
one spin species.  Thus our results are general, and can be
translated to the positive-$U$ case \cite{emery}.



For large negative $U$ the fermions form tightly bound pairs that
behave 
as hard-core bosons \cite{emery}.  These
bosons are prevented from fully condensing in 1D due to quantum
fluctuations. 
They form a ``quasi-condensate", with pair correlations that decay
as a power-law that, in some regime of parameters (large $|U|>4t$)
can dominate over the single-fermion correlations at large
distances \cite{Yang}. In the polarized
case, the ground-state of this system is 
the 1D version of the 
Fulde-Ferrell-Larkin-Ovchinnikov (FFLO) superfluid
\cite{ff,lo,Yang}, in which the pairs forming the quasi-condensate
have nonzero center of mass momentum $\pm Q$ with 
$Q=k_{F\uparrow}-k_{F\downarrow}$, where $k_{F\sigma}$ is the
Fermi momentum of the fermions with spin $\sigma$. This was
confirmed numerically in Ref.\cite{ourFFLO}, and subsequent studies \cite{ueda,batrouni}).


The Luttinger-liquid and FFLO aspects of this system 
can be heuristically understood as follows:  At large negative
$U$, the spin-polarized ground state consists of empty sites (0's),
sites occupied by pairs (2's) and excess up fermions
($\uparrow$'s), with sites singly-occupied by $\downarrow$'s being
only ``virtual'' states. The density of excess $\uparrow$'s is
$Q/\pi$.  An $\uparrow$ 
exchanges positions with the 0's and 2's with hopping $t$ and thus
moves with bandwidth $4t$.  At half-filling, the background the
$\uparrow$ moves through is half 0's and half 2's, so the relative
motion moves spin but no density on average: this collective mode
is then purely a spinon.  But if we move away from half-filling,
the number of 0's differs from the number of 2's, so when an
$\uparrow$ moves, it on average moves some density as well as
spin: this light (bandwidth $4t$) mode of 
the Luttinger liquid is then not purely spin, but instead is a
particular linear combination of spin and charge (we will call
this light mode ``spinon-like''). In the limits of nearly complete
polarization or either zero or complete filling, the $\uparrow$'s
become just regular fermions carrying the full charge and spin.
This scenario has been confirmed numerically in Ref.
\cite{kollath1,kollath2,kollath3}, by looking at the real-time
evolution of spin and charge distributions.

At large negative $U$ the 2's do not move freely past the 0's;
this exchange happens via a virtual intermediate unpaired state
with energy $|U|$, resulting in effective hopping
$t_{eff}=-2t^2/U$. Thus this motion of 2's relative to 0's
constitutes the heavy ``holon-like'' mode of the Luttinger liquid
with a smaller bandwidth. Also, when a 2 moves past an $\uparrow$,
the ground state has a sign change.
This means the wavefunction of the quasi-condensate of bosonic 2's
has a node at each $\uparrow$.  If these nodes were
equally-spaced, this would be an FFLO standing-wave condensate
with momentum $\pm Q$.  However, the $\uparrow$'s actually form a
1D Luttinger liquid with divergent position fluctuations, so the
momentum distribution of the pairs instead has a power-law
divergence at $\pm Q$; this 1D partially spin-polarized superfluid
state should perhaps be termed ``quasi-FFLO".

The Hamiltonian (\ref{one}) can be solved exactly by means of the
Bethe Ansatz \cite{lieb and wu,woynarovich,woynarovich and penc},
and the dispersion of the elementary excitations can be obtained
\cite{schulz,essler,andrei's review}. However, the actual Green's
functions and spectral properties can only be calculated in
certain limits \cite{penc}, and numerical methods have been
crucial to fill in the blanks and compare to experiments
\cite{ED,jeckelmann}.  In the following, we use the time-dependent
extension of the Density Matrix Renormalization Group (tDMRG)
\cite{tdDMRG1,tdDMRG2} method to obtain estimates for various
Green functions in real-time and real-space with unprecedented
accuracy \cite{specs}.
To extract the dynamical
response of the system, we calculate the correlators
$G(x-x',t'-t)=i\langle O(x',t')O^\dagger(x,t) \rangle$, where $O$
is an operator of
interest. 
Fourier transforming 
then yields the corresponding spectral weights as functions of
momentum and frequency \cite{specs,tdDMRG1,White and Affleck}:
\begin{eqnarray}
I(k,\omega)=\sum_{n} |\langle \psi_{n}|O_k|\psi_0 \rangle|^2
\delta(\omega-E_{n}+E_0) ~,
\end{eqnarray}
where $E_0$ is the ground state energy, and the sum runs over all
the eigenstates of the system, with energy $E_n$. All the results will be
plotted using a log-scale for the intensity, with several orders of magnitude between the intensities of the
weakest and strongest features. At very small scales, some ripples or
oscillations appear as a consequence of the numerical Fourier transform, and the commensuration of the lattice. These effects get amplified near zero momentum and frequency.

\begin{centering}
\begin{figure}
\epsfig {file=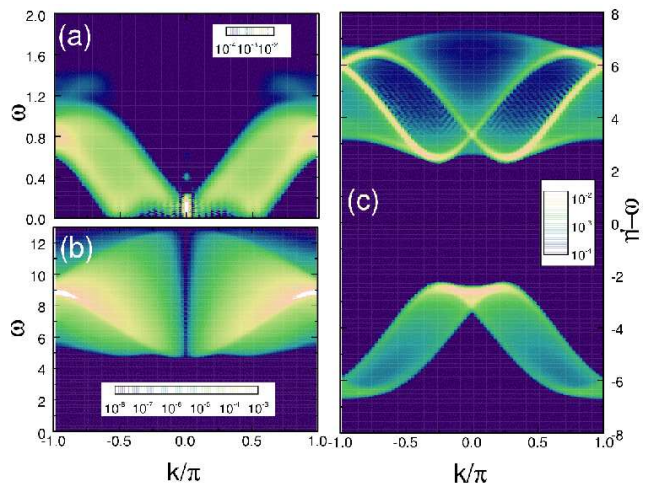,width=65mm,angle=-90} \caption{
(color online)
Dynamical structure factors of the (a) ``charge" density
$n(k,\omega)$, and (b) spin $S_z(k,\omega)$ for an unpolarized,
quarter-filled Hubbard chain with $U=-8t$. (c) Spectral weights
for adding ($\omega>\mu$) or removing ($\omega<\mu$) a fermion for
the same system; $\mu$ is the chemical potential. Frequencies are
in units of the hopping $t=1$.  The colors are set by the
logarithm of the spectral intensity.
} \label{fig1}
\end{figure}
\end{centering}

In Fig. \ref {fig1}(a) and (b) we show the dynamic structure factor
for the charge and spin densities, respectively, for an
unpolarized Hubbard chain at quarter-filling (in this paper we always use $L=80$ and
$U=-8t$). 
The charge excitations
display gapless modes at momenta $k=0$ and $k=\pm 2k_F=\pm\pi/2$,
and a continuum ranging from $\omega=0$ to
$\omega \cong t=4t_{eff}$. This spectrum is formed primarily by 
holon-antiholon excitations. 
It is qualitatively similar to the particle-hole spectrum of the
corresponding non-interacting system, but with a reduced
bandwidth.  However, this system is a superfluid with a spin gap
of $\cong 5t$, as is seen in the spectral weight of the spin (Fig.
1b); this is the energy ``cost" of breaking a Cooper pair. The
spinon has band-width $\cong 4t$, and the spectral weight of $S_z$
vanishes strongly as $k \rightarrow 0$, since the total spin is
conserved and the matrix element for making spin excitations thus
vanishes at zero momentum. 

The single-particle spectral weight for the quarter-filled,
unpolarized system is shown in Fig. \ref{fig1}(c), where we plot
the imaginary part of the one-particle Green's function. The upper
and lower features, for positive and negative frequencies,
correspond to the inverse photoemission (IPES) and photoemission
(PES) spectra,
resulting from adding or removing 
a fermion, 
respectively.
 We have shifted the energies relative to the chemical potential
$\mu=[E_0(N+1)-E_0(N-1)]/2$, which lies in the center of the gap
for this unpolarized system.  This gap is a manifestation of the
spin gap due to Cooper pairing: the ground state is a total spin
singlet with all fermions paired.  The added fermion has no
``partner'' to pair with, while removing a fermion requires
breaking an existing pair, so both processes are gapped.

Again, we can heuristically understand many features of these
spectra using the large-negative-$U$ description discussed above.
The unpolarized ground state is a quasi-condensate of 2's that
form a Luttinger liquid of repulsively interacting bosons.  An
added $\uparrow$ forms a spinon and much of its spectral weight
thus follows a spinon dispersion with bandwidth $4t$. Since the
wavefunction changes sign when the $\uparrow$ exchanges position
with a 2, the lowest-energy spinon states are at the momenta $\pm
\pi/4$ set by the density of the 2's. However, the added fermion
may also excite holon modes, and a careful look at the upper part
of Fig. \ref{fig1}(c) reveals a continuum, with a weaker feature at
the lower edge of the continuum which has a holon-like dispersion.
This continuum arises when part of the added momentum is used to
excite holon modes of the quasi-condensate.

Removing a fermion requires breaking a pair (a ``2''), and
this process apparently couples more strongly to the holon degrees
of freedom, as can be seen by the flatter dispersion of the strong
part of the spectrum at low momentum in the bottom part of Fig.
\ref{fig1}(c).  However, at higher momentum, this PES spectrum,
although much weaker, has a continuum with a mostly spinon-like
dispersion. Here the process apparently removes a low-momentum
pair from the quasi-condensate and makes a spinon, with the spinon
taking most of the momentum.  At half-filling,
 there is particle-hole symmetry and the
PES and IPES spectra are thus equivalent, both containing strong
spinon and holon signals.\cite{jeckelmann}

\begin{centering}
\begin{figure}
\epsfig {file=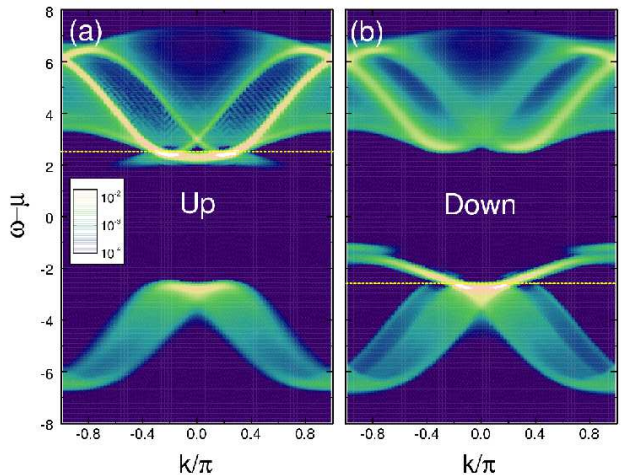,width=65mm,angle=-90} \caption{
(color online)
``Photoemission'' spectra for quarter-filled spin-polarized
Hubbard chain, with $U=-8t$, $N_\uparrow=24$, $N_\downarrow=16$.
The energy scale has been chosen relative to the average chemical
potential $\mu=(\mu_\uparrow+\mu_\downarrow)/2$.
The Fermi levels 
for each spin species are indicated by the horizontal lines (see
text).  
} \label{fig2}
\end{figure}
\end{centering}

We now turn our attention to the single-fermion spectrum in the
polarized case, shown in Fig. \ref{fig2}, where we took
$N_\uparrow-N_\downarrow=8$.  Since the system is no longer
symmetric under time-reversal, the spectral functions for the up
and down fermions are different. Correspondingly, we can determine
the 
chemical potential of each species:
$\mu_\sigma=[E(N_\sigma+1)-E(N_\sigma-1)]/2$.  The average
chemical potential is 
$\mu=(\mu_\uparrow+\mu_\downarrow)/2$, while the effective Zeeman
field is $h=(\mu_\uparrow-\mu_\downarrow)/2$.
We find it instructive to plot the spectra with energies relative to 
$\mu$. 
Note that $\mu$ is still in the ``pairing'' gap, but now
$\mu_{\uparrow}$ is in the band above the gap, while
$\mu_{\downarrow}$ is in the band below the gap.

At this fairly large $|U|$, we can describe this system as a
quasi-condensate of bosonic 2's with density
$k_{F\downarrow}/\pi=1/5$ and momentum $\pm Q=\pm\pi/10$, plus a
density $Q/\pi=1/10$ of excess unpaired $\uparrow$'s.  The 2's are
bound pairs and sit below the gap and just below
$\mu_{\downarrow}$.  In the PES spectrum one can remove a fermion
of either spin from one of these singlet pairs; these are the
strong low-momentum features near $\omega-\mu\cong -3$.  The
weaker bands dispersing strongly to lower energy from these
features arise from removing one member of a pair and leaving the
other member in a spinon-like state.

The excess unpaired $\uparrow$'s lie at energy just below
$\mu_{\uparrow}$ and can be seen there in the spin-up PES
spectrum.  The wavefunctions of the $\uparrow$'s change sign on
passing each 2; as a result the lowest energy states of the
corresponding spinon-like modes are at $\pm k_{F\downarrow}$; it is near these
momenta where the up PES intensity is largest.  The strongest
bands in the up spectrum cross $\mu_{\uparrow}$ at $\pm
k_{F\uparrow}$, just as in the noninteracting system.  But one can
also see weaker bands crossing $\mu_{\uparrow}$ at $\pm
(2k_{F\downarrow}-k_{F\uparrow})$, which correspond to
three-particle excitations in the noninteracting system.  At low
momentum and energies below $\mu_{\uparrow}$ there is a fairly
flat holon-like dispersion of the up spectral weight; presumably
here the excitation also transfers some momentum to the (heavy)
pairs.

In the spin-down IPES spectrum, there is a heavy holon-like band
at energies just above $\mu_{\downarrow}$.  This arises from
adding a down fermion that pairs with one of the excess up
fermions with momentum $|k|\leq k_{F\uparrow}$, resulting in a
pair (a 2) which carries most of the added momentum.  This feature
in the IPES is strong only for $|k|\geq k_{F\downarrow}$, since
the down-spin states at lower momentum than this are already
occupied.  This band continues to the PES spectrum below
$\mu_{\downarrow}$, crossing the chemical potential at $k=\pm
k_{F\downarrow}$, as in the noninteracting system.  At the zone
boundary this holon-like band splits in to two faint features at
$(\omega-\mu)$ near -1 and -1.5, for reasons we do not yet
understand.  In the PES spectrum there are also weaker features
approaching $\mu_{\downarrow}$ at momenta $\pm
(2k_{F\uparrow}-k_{F\downarrow})$, which again correspond to
three-particle excitations in the noninteracting system.

In the IPES spectrum above the gap, we can see that for both spins
the added particle can excite a continuum of states with both
spinon-like and holon-like dispersions.  The sharpest feature is a
spinon band, which is substantially sharper 
in the up IPES than in the down. The general appearance of this
part of the IPES spectral weight is similar to that of the
unpolarized case in Fig.1(c).

\begin{centering}
\begin{figure}
\epsfig {file=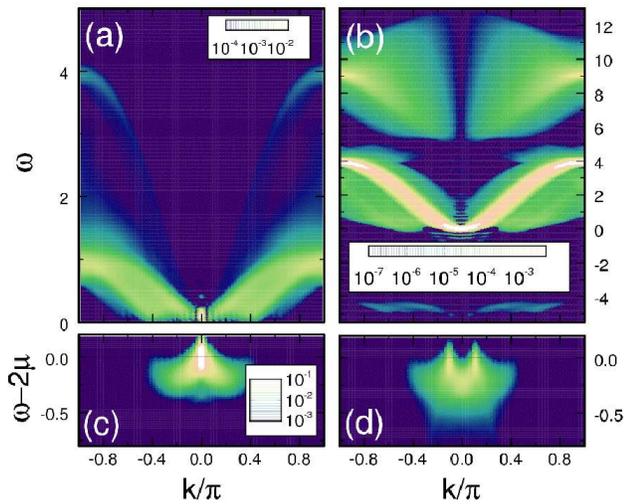,width=85mm} \caption{
(color online)
Dynamical structure factors of the (a) ``charge" density
$n(k,\omega)$, and (b) spin lowering operator $S_-(k,\omega)$ for the polarized,
quarter-filled Hubbard chain of Fig. 2.
(c) 
Spectral weight for removing a pair from the unpolarized system, and
(d) the polarized system. 
In (c) and (d) the energy scale is
relative to the chemical potential of a pair $2\mu$. } \label{fig3}
\end{figure}
\end{centering}

In order to examine the quasi-condensate and the pair-like
excitations, we calculated the spectral weight of removing a pair
$b_i=c_{i\uparrow}c_{i\downarrow}$, shown
in Fig. \ref{fig3}(c) and (d). For the unpolarized case, we see a large
concentration of spectral weight at zero energy and zero momentum
in the PES, due to the zero-momentum quasi-condensate.
In the polarized case, the zero-energy spectral weight of this quasi-condensate 
splits into two features at $k=\pm Q=\pm (k_{F\uparrow} -
k_{F\downarrow})$, as expected for this FFLO-like state.

In Fig. \ref{fig3}(a) we show the dynamical structure factor for
the density operator in the polarized case.  One noteworthy 
difference from the unpolarized case is the appearance of a
weak spinon-like feature.  This occurs because in this polarized system away from half-filling, the spinon-like mode is no longer purely spin so couples to the density.

Since
time-reversal symmetry is broken by the spin polarization, the response functions for the
spin operators $S_z$, $S_+$, and $S_-$ are now all different. 
The structure factor for $S_z$ (not shown) exhibits, besides the
excitations across the gap present in the unpolarized system (Fig. 1(b)), a
gapless band that originates from spin excitations within the bands that cross the
Fermi surfaces. The operator $S_+$ flips spins up, breaking pairs, 
so its spectrum only shows excitations across the gap. 
The richest of these spin dynamical functions is
$S_-(k,\omega)$, shown in Fig.\ref{fig3}(b). The action of the operator $S_-$
on the ground state can cause three possible outcomes that each occupy a separate energy window: 1) it
can break a pair; 2) it can flip an unpaired
$\uparrow$ to an unpaired $\downarrow$, making a gapless
spin fluctuation; or 3) the flipped spin can pair with another unpaired
$\uparrow$ and be absorbed by the quasi-condensate.  We find that the
spectral weight for this last process is very weak, but detectable; it is visible just below energy -4 in Fig. 3b.
We believe this very small weight is due to the product of two small factors:
the low probability (due to fermionic antisymmetry) that
two unpaired $\uparrow$'s are on adjacent lattice sites before one of them is flipped down, and the low
overlap between the resulting state after flipping, and the ground state with a bound pair, since the latter mostly consists of
doubly-occupied sites.

To summarize, we have reported and discussed the rich features of the particle, pair, spin and density
spectral weights for the quasi-FFLO superfluid ground state of a partially spin-polarized fermionic Hubbard chain with attractive interactions. We have found that a rigorous treatment, particularly to
describe properties involving excitations, should still rely
on the Luttinger liquid picture.

\acknowledgements

AEF is grateful to F. Heidrich-Meisner, M. Troyer, M.P.A. Fisher,
and C. Nayak for useful and stimulating discussions. This work was
partially supported (DAH) under ARO Award W911NF-07-1-0464 with
funds from the DARPA OLE Program.


\begin{thebibliography}{99}
\bibitem{specs} We typically used 400 DMRG states, keeping the
truncation error $<10^{-7}$, and a 3rd-order Suzuki-Trotter
expansion of the evolution operator with time-step $\tau=0.02$
or less (in units where the hopping is $t=1$).
We calculated the time evolution up to time $30$, and
Fourier transformed using a cosine window.
\end{thebibliography}
\end{document}